\documentstyle[12pt]{article}
\begin{document}
\begin{flushright}
{\bf hep-th/9911133}\\
{\bf UCVFC-DF/17-99}
\end{flushright}
\vskip 1cm
\begin{center}
{\LARGE Knot Invariants from Classical Field Theories}

\vskip 12pt

{\Large Lorenzo Leal}\footnote{Talk delivered at the Spanish Relativity
Meeting, Bilbao, Spain, 1999}\\

{\it Departamento de F\'{\i}sica, Facultad de Ciencias, \\
Universidad Central de Venezuela, AP 47270,\\
Caracas 1041-A, Venezuela}\\{\em Email: {\tt lleal@tierra.ciens.ucv.ve}}\\   
\end{center}
\abstract{
We consider the Non-Abelian Chern-Simons term coupled to external
particles, in a gauge and diffeomorphism invariant form. The
classical equations of motion are perturbativelly studied, and the on-shell
action is shown to produce knot-invariants associated with the sources. The
first contributions are explicitely calculated, and the corresponding
knot-invariants are recognized. We conclude that the interplay between Knot
Theory and Topological Field Theories is manifested not only at the quantum
level, but in a classical context as well.}

\newpage

\section{Introduction}

The study of quantum Chern-Simons theory (C.S.T.) and its relationship with
Knot Theory (K.T.) is an active field of research both in Mathematics and
Theoretical Physics \cite{labastida}. The bridge between field and knot theories
was stablished when E.Witten discovered that the vacuum expectation value of
the Wilson Loop (W.L.) yields a knot invariant closely related with the
Jones polinomial \cite{witten,jones}.
From a physical point of view, this relationship has interesting
applications. For instance, the knot invariants obtained from a perturbative
expansion of the W.L. average \cite{guadagnini} provide solutions to  the
 constraints of Ashtekar Quantum Gravity \cite{ashtekar} in the Loop
Representation \cite{gambini,rovelli}.
The purpose of this work is to show that the interplay between C.S.T. and
K.T. is manifested also at the classical level. To this end, we shall
develop a perturbative study of the classsical equations of motion of the
C.S.T. coupled to external point particles. As we shall see, this study
leads to the obtention of link invariants associated to the world lines of
the particles, inasmuch the calculation of the expectation value of the W.L.
does in the quantum theory. This approach was explored by the author for the
abelian case several years ago \cite{leal}.

\section{Classical Chern-Simons Theory}

\subsection{Preliminaries}

The SU(N) action is given by

\begin{equation}
S_{CS}=-k\int d^3x\, \epsilon^{\mu\nu\rho}\,
tr(A_\mu\partial_\nu A_\rho +  \frac{2}{3}A_\mu A_\nu A_\rho)
\label{eq:scs}
\end{equation}

\noindent with \begin{math} A_\mu=A_\mu^aT^a \end{math}. The $N^2-1$ generators 
$T^a$ are chosen so that
$trT^aT^b=-\frac{1}{2}\delta^{ab}$ and
$[T^a,T^b]=\Gamma^{abc}T^c$, $\Gamma^{abc}$ being the (completelly 
antisymmetric) structure constants.
We shall add an interaction term

\begin{equation}
S_I=\int d^3x \,tr(A_\mu J^\mu)
\label{eq:si}
\end{equation}

\noindent with

\begin{equation}
J^\mu (x)=\sum_{i=1}^n \int d\tau
\delta^{3}(x-z(\tau))\dot{z}_{(i)}^{\mu}(\tau)I_{(i)}(\tau)
\label{eq:jmu}
\end{equation}

\noindent In this expression, $z^\mu_{(i)}(\tau)$ denotes the wold line of
 the $i-th$ particle, which
acts as an externally prescribed source for the C.S. field. Indeed, since we
are going to take these world lines as closed curves in 2+1 dimensions, the
analogy with point particles holds only formally.
The algebra valued object $I_{(i)}(\tau)=I_{(i)}^a(\tau)T^a$ could be
 interpreted as the color charge
carried by the $i-th$ particle. The current $J^\mu(x)$ was introduced by
Wong~\cite{wong} to deal
with point particles interacting with the Yang-Mills field.
Varying $S_{CS}+S_I$ w.r.t. $A_\mu$ produces the field equations

\begin{equation}
k\epsilon^{\mu\nu\rho}\,F_{\nu\rho}=J^\mu
\label{eq:movimiento}
\end{equation}

\noindent where
$F_{\mu\nu}=\partial_{\mu}A_{\nu}-\partial_{\nu}A_{\mu}+[A_{\mu},A_{\nu}]$.
 Consistence of Eq.(4) leads to

\begin{equation}
D_{\mu} J^{\mu} \equiv \partial_\mu J^\mu +[A_\mu ,J^\mu]=0
\label{eq:Dmu}
\end{equation}

\noindent Suppose now that one is able to solve
 Eqs.(\ref{eq:movimiento}),(\ref{eq:Dmu}) to get $A_\mu=A_\mu(J)$, and to
calculate the on shell action $S_{OS}=S[A(J)]$ . Since both $S_{CS}$ and $S_I $
 are topological terms, it is immediate to see that $S_{OS}$ is metric
 independent too. Moreover, since $S_{OS}$ depends only on the curves
$\dot{z}_{(i)}^{\mu}(\tau)$ , we have to
conclude that it is a link-invariant associated to the curves.
Of course, one cannot solve the non-linear equations
 (\ref{eq:movimiento}),(\ref{eq:Dmu}) exactly. Therefore,
we shall adopt a perturbative scheme, with ${\Lambda}=k^{-1}$ being the
 expansion parameter. We shall get

\begin{equation}
S_{OS}(J) = \sum_{p=0}^\infty\,\Lambda^p S^{(p)}(J)
\label{eq:sos}
\end{equation}

\noindent and, since $S_{OS}(J)$ is a link-invariant for all $\Lambda$ , the
 same will be true for each contribution $S^{(p)}$.

\subsection{Perturbative Solution}
Next we sketch how the perturbative procedure goes on, and calculate
$S_{OS}$ explicitly up to first order in $\Lambda$.
It can be seen that the consistence equation (\ref{eq:Dmu}) leads to

\begin{equation}
\frac{dI^a_{(i)}(\tau)}{d\tau}+R^{ac}_{(i)}(\tau)I^c_{(i)}(\tau)=0
\label{eq:paraltrans}
\end{equation}

\noindent with

\begin{equation}
R^{ac}_{(i)}(\tau)\equiv\Gamma^{abc}\dot{z}_{(i)}^{\mu}(\tau)B_{\mu}^{b}(z_{(i)}(\tau))
\label{eq:defR}
\end{equation}

\noindent and $B_{\mu}\equiv\Lambda^{-1}A_\mu$ . Eq. (\ref{eq:paraltrans})is
 solved by

\begin{equation}
I_{(i)}(\tau)= Texp\left[ -\Lambda\int_{0}^{\tau}d\tau'R_{(i)}(\tau')
\right]I_{(i)}(0)
\label{eq:Texp}
\end{equation}

\noindent where $T$ denotes the path ordered exponential along the curve
$z_{(i)}^{\mu}(\tau)$. Putting
Eq.(\ref{eq:Texp}) into Eq.(\ref{eq:movimiento}) one obtains, after developing
 the ordered exponential

\begin{eqnarray}
& &\epsilon^{\mu\nu\rho}(\partial_{\nu}B_{\rho}^a-\partial_{\rho}B_{\nu}^a)=
\nonumber\\
& & \qquad \qquad \sum  _{i=1}^n \oint dz_{(i)}^\mu {\delta}^3 (x-z_{(i)})
 I_{(i)}^a(0) +\Lambda \epsilon^{\mu\nu\rho}\, \Gamma^{abc}\, B_\nu^b
 B_\rho^c\nonumber\\
& & \qquad \qquad -\Lambda \sum  _{i=1}^n \oint dz_{(i)}^\mu
 {\delta}^3 (x-z_{(i)}) \int_0^{z} dz ^{\mu_1}_{(i)1} R^{aa_1}_{\mu_1}
(z_{(i)1})I_{(i)}^{a_1} (0)\nonumber\\
& & \qquad \qquad +\Lambda^2  \sum  _{i=1}^n \oint dz_{(i)}^\mu
 {\delta}^3 (x-z_{(i)}) \int_0^{z} dz ^{\mu_1}_{(i)1} \int_0^{z_1}
 dz ^{\mu_2}_{(i)2}R^{aa_1}_{\mu_1}(z_{(i)1})R^{a_1 a_2}_{\mu_2}(z_{(i)2})
I_{(i)}^{a_2} (0)\nonumber\\
& & \qquad \qquad \qquad \vdots 
\label{eq:maestra}
\end{eqnarray}

\noindent To solve this equation perturbatively, we introduce the power
 expansion

\begin{equation}
B_\mu^a =\sum_{p=0}^\infty \Lambda^p \stackrel{(p)}{B_\mu ^a}
\label{eq:Bpowers}
\end{equation}

\noindent  into Eq.(\ref{eq:maestra}). This yields, for the $0th$ order

\begin{equation}
\epsilon^{\mu\nu\rho}(\partial_{\nu} \stackrel{(0)}{B_\rho ^a}-\partial_{\rho}
 \stackrel{(0)}{B_\nu ^a})=\sum  _{i=1}^n \oint dz_{(i)}^\mu {\delta}^3
 (x-z_{(i)})I_{(i)}^a(0)
\label{eq:Omovimiento}
\end{equation}

\noindent The first order equation is given by

\begin{eqnarray}
\epsilon^{\mu\nu\rho}(\partial_{\nu}\stackrel{(1)}{B_\rho ^a}
 -\partial_{\rho}\stackrel{(1)}{B_\rho ^a}) 
&=& \epsilon^{\mu\nu\rho}\, \Gamma^{abc}\, 
\stackrel{(0)}{B_\nu ^b}\,\stackrel{(0)}{B_\rho ^c}\nonumber\\
&-& \sum  _{i=1}^n \oint dz_{(i)}^\mu {\delta}^3 (x-z_{(i)})
 \int_0^{z} dz ^{\mu_1}_{(i)1} \stackrel{(0)}{R^{aa_1}_{\mu_1}}
(z_{(i)1})I_{(i)}^{a_1} (0) 
\label{eq:1movimiento}
\end{eqnarray}

\noindent For the sake of brevity we omit the $pth$ equation. Its general
 structure is given by

\begin{equation}
\epsilon^{\mu\nu\rho}\partial_{\nu} \stackrel{(p)}{B_\rho ^a} = 
\stackrel{(p)}{J^{\mu a}}
\label{eq:ampere}
\end{equation}

\noindent where $ \stackrel{(p)}{J^{\mu}}$ depends on $\stackrel{(q)}{B_\mu}$
 , with $q<p$. This feature allows one to solve Eq. (\ref{eq:maestra})
order by order in a straightforward manner. In fact, $\stackrel{(p)}{B}$ is
 given by

\begin{equation}
\stackrel{(p)}{B_\alpha ^a}(x)= -\frac{1}{4\pi} \int d^3 x'
 \epsilon_{\alpha\beta\gamma} \stackrel{(p)}{J^{\beta\gamma}}
(x')\frac{(x-x')^\gamma}{|x-x'|^3}
\label{eq:biotsavart}
\end{equation}

\noindent plus the gradient of an arbitrary function, which can be set equal
 to zero. This amounts to chosing the gauge $\partial^{\nu} B_\nu ^a = 0$.
 Observe that for the first time, a metric (the Euclidean one) appears into
 the discussion. This will not spoil the diffeomorphism invariance of the
 on-shell action, because the latter is not sensible to metric choices. It
 could be said that chosing a metric to solve the equations amounts to fixing
 a gauge, or a "geometry", that will not break the topological invariance of
 the "geometry-independent" on-shell action.
As before, there are consistence requirements which must be studied: from
Eq.(\ref{eq:ampere}) it is immediate that $\stackrel{(p)}{J^{\mu a}}$
 must be conserved. This is clearly fulfiled in the $0th$ order case, but as
 we shall discuss later, it is by no means trivial for higher orders.
We now turn to the on-shell action. We set

\begin{eqnarray}
\mathcal{S} & \equiv & - \frac{2}{\Lambda} S_{OS} \nonumber\\
& =& \int d^3x\, \epsilon^{\mu\nu\rho}\,
(B^a_\mu\partial_\nu B^a_\rho +  \frac{2}{3} \Lambda\, B^a_\mu B^b_\nu
 B^c_\rho\,\Gamma^{abc})_{B=B(J)}
\label{eq:sdebdej}
\end{eqnarray}

\noindent where we have used the equation of motion Eq.(\ref{eq:movimiento}) to
 eliminate $J^{\mu}$ in terms of $B_{\mu}$. Using equations
 (\ref{eq:Omovimiento}) and (\ref{eq:biotsavart}), we have, up to $0th$ order

\begin{eqnarray}
\stackrel{(0)}{\mathcal{S}} & = & \int d^3x\, \epsilon^{\mu\nu\rho}\,
(\stackrel{(0)}{B^a_\mu}\partial_\nu \stackrel{(0)}{B^a_\rho} )_{B=B(J)}
\nonumber\\
& = & \frac{1}{4} \sum_{i,j} I^a_i (0) I^a_j(0) {\mathcal{L}}(i,j) 
\label{eq:0sos}
\end{eqnarray}

\noindent where

\begin{equation}
{\mathcal L}(i,j) = \frac{1}{4\pi} \int_i dz^\mu \int_j dz'^\rho \frac{(z-z')
^\nu}{|z-z'|^3} \,\epsilon_{\mu\nu\rho}
\label{eq:linking}
\end{equation}

\noindent is the Gauss Linking Number of the curves $i,j$, which is a link
 invariant. (There are some subtleties about the case $i=j$, whose discussion
 is out the scope of this paper~\cite{guadagnini}). 
To obtain the first order contribution $\stackrel{(1)}{\cal{S}}$, we have to
 study the consistence of Eq. (\ref{eq:1movimiento}). Taking the divergence on
 both sides, and after some calculations, one is lead to the condition

\begin{equation}
\sum_{i,j} \Gamma^{abc} I^b_i(0) I^c_j(0)\,{\mathcal L}(i,j)\,\delta
^3(x-z_{(i)}(0)) = 0
\label{eq:condition}
\end{equation}

\noindent where $z_{(i)}(0)$ is the starting point of the curve $i$ . From the
 last equation we see that a sufficient condition for Eq. (\ref{eq:1movimiento})
 being consistent is that ${\mathcal L}(i,j) = 0 \; \forall\; i,j $. A careful
 look at Eq.(\ref{eq:condition}) reveals that if the curves do not intersect
 each other, the former is indeed a necesary condition too. Under these
conditions, Eq.(\ref{eq:biotsavart}) gives the solution of Eq.
(\ref{eq:1movimiento}), which when substituted into Eq.(\ref{eq:sdebdej})
 yields the first order contribution to $\cal{S}$

\begin{equation}
\stackrel{(1)}{\cal{S}} = \sum_{i,j,k} G_{i,j,k} M(i,j,k)
\label{eq:gporm}
\end{equation}

\noindent where $G_{i,j,k}$ is a group theoretical factor and

\begin{eqnarray}
 M(i,j,k) &=&\int d^3x\epsilon^{\mu\nu\rho}C_{(i)\mu}(x)C_{(j)\nu}(x)
C_{(k)\rho}(x)\nonumber\\
 &+&\int d^3x\int d^3y
\;\big[\,T^{\mu\nu}_{(i)}(x,y)C_{(j)\mu}(x)C_{(k)\nu}(y)\,+\;cyc. perm. of
\;i,j,k\big]
\label{eq:mijk}
\end{eqnarray}

\noindent with

\begin{equation}
C_{(j)\mu}(x)\equiv \frac{1}{8\pi}\oint_i dz^\gamma \frac{(x-z)^\beta}{|x-z|^3}
\,\epsilon_{\mu\beta\gamma}
\label{eq:alexdual}
\end{equation}

\noindent and

\begin{equation}
T^{\mu\nu}_{(i)}(x,y)\equiv\oint_idz^{\mu}\int_o^z dz'^{\nu}\delta^3 (z-x)\
delta^3 (z'-y)
\label{eq:txy}
\end{equation}

\noindent The quantity defined in Eq. (\ref{eq:mijk}) is the Triple Milnor's
 Linking Number \cite{milnor,monastri,reshetkin} associated to the curves
 $i,j,k,$ which is known to be a link-invariant, provided the Gauss Linking
 Numbers vanish for every $i,j$. Hence, we obtain the following nice feature:
 the consistence condition for solving the first order equation of motion, is
preciselly the condition which allows the first order contribution to the
on-shell action to coincide with a non-trivial link-invariant. As an application
 of this invariant, we recall that the Borromean Rings have non-vanishing
 $ M(i,j,k)$ while their Gauss Linking Number vanish~\cite{monastri}`.

\section{Discussion}

We have presented results supporting the following claim: Classical (not
only Quantum) Topological Field Theories are an interesting tool for the study
 of link invariants. Concretely, we found that the on shell action of the C.S.T.
coupled with string-like sources, admits a perturbative expansion in powers
of the inverse coupling constant, whose first and second order contributions
 give respectively the Gauss Linking Number and the Triple Milnor's Linking
 Number. We want to underline that our proposal and results are
completely independent of quantum considerations.

\noindent The perturbative scheme we have reported could certainly be carried
 out further. We conjecture that as far as the links allowed do not intersect
 each other, the corresponding higher order contributions to the on-shell action
will yield Milnor's higher-order linking coeficients. This should be compared
 with similar results of reference \cite{reshetkin}, obtained within the
 quantum framework. It would also be interesting to study the case where the
 curves intersect. These and other related issues are under work.
\section{Acknowledgments}
 I wish to thank to the Organizers of the ERE-99 and to the City of Bilbao 
for their warm hospitality. This work was supported by CDCH, Universidad
Central de Venezuela. 

\vspace*{-9pt}

\end{document}